# Differential femtosecond coherent Stokes and anti-Stokes Raman spectroscopy

Takuro Ideguchi,[1] Simon Holzner,[1] Ming Yan,[1,2] Guy Guelachvili,[3] Theodor W. Hänsch,[1,2] and Nathalie Picqué [1,2,3,] ∗

[1] Max Planck Institut für Quantenoptik, Hans-Kopfermann-Strasse 1, 85748 Garching, Germany

[2] Fakultät für Physik, Ludwig-Maximilians-Universität München, Schellingstrasse 4/III, 80799 München, Germany

[3] Institut des Sciences Moléculaires d'Orsay, CNRS, Bâtiment 350, Université Paris-Sud, 91405 Orsay, France

∗Corresponding author: nathalie.picque@mpq.mpg.de

**Abstract:** We demonstrate a novel technique of coherent Raman spectroscopy with a femtosecond laser. We apply to a molecular sample a sequence of pairs of ultrashort excitation and probe pulses, with a linearly increasing time delay between the two pulses from one pair to the next. We measure, as a function of the delay, the intensity modulation in the signal resulting from the differential detection of the Stokes and anti-Stokes radiations generated at the sample. The Fourier transform of such time-domain signal reveals the spectrum of the excited vibrational Raman transitions. The experimental proof-of-principle demonstrates high resolution, broad spectral span and suppression of the non-resonant background, as well as sensitivity enhancement due to the differential detection.

Over the four last decades, coherent anti-Stokes Raman spectroscopy (CARS) has evolved into a powerful tool for chemical sensing and imaging. A broad high-resolution vibrational spectrum can be obtained by multiplex –or multichannel- CARS techniques [1–9], which take advantage of femtosecond lasers. Such noninvasive techniques provide a good chemical selectivity in complex samples. In this letter, we report on a new technique of coherent Raman spectroscopy that allows for the measurement of a broad (>1000 cm$^{-1}$) high resolution (4 cm$^{-1}$) spectrum with a single differential photodetector system. We exploit time-domain impulsive Raman spectroscopy and simultaneously record the Stokes (red-shifted) and the anti-Stokes (blue-shifted) signals for differential detection and improved signal-to-noise ratio. Based on an ultrashort pulse laser and a scanning Michelson interferometer, our technique could benefit the various implementations of impulsive coherent Raman spectroscopy.

Impulsive coherent Raman spectroscopy is a well known time-domain technique [10] that produces broadband high-resolution spectra with ultrashort pulse lasers. A sequence of pairs of excitation and probe pulses, separated by a time delay that increases linearly from pulse pair to pulse pair (Fig.1), hit the sample. An excitation pulse excites low-lying molecular vibrational energy levels in a two-photon Raman process. The coherent vibrations induce a modulation of the refractive index of the sample at the vibrational frequencies. Assuming that the probe pulse is short compared to the molecular vibration periods, when it arrives onto the sample after half a period, the vibrational motion is damped and the pulse experiences an anti-Stokes (blue) shift. When it arrives after a full modulation period, the vibration amplitude is increased and the probe pulse is Stokes (red) -shifted. In previous reports of impulsive coherent Raman spectroscopy, the modulation of the intensity of the blue-shifted light after an optical shortpass filter has been monitored as a function of the time delay between the excitation and probe pulses. The resulting time-domain signal, the interferogram, is Fourier transformed to reveal the spectrum of all the excited vibrational transitions of the sample. The varying time delay in the pulse pairs has been achieved by several means: with a variable mechanical delay line (e.g. a Michelson interferometer) [4–7,11,12], with a spatial light modulator [8] or more recently with a dual-comb system [9]. All techniques follow the same physical principle and share several distinguishing features including a broad spectral span determined by the laser pulse duration only, a high spectral resolution determined by the measurement time only, the entire suppression of the non-resonant background and the use of a single photodetector. The newly demonstrated implementation of dual-comb impulsive coherent Raman spectroscopy without moving parts [9] additionally reports recording times for an interferogram on the microsecond scale, though the refresh rate of successive interferograms is currently hundreds-fold longer. In this letter, we explore a way to improve the sensitivity of the measurements in impulsive coherent Raman spectroscopy.

In coherent Raman spectroscopy, detection of the Stokes-shifted light, often called coherent Stokes Raman spectroscopy (CSRS), is not widespread. The anti-Stokes light is indeed free of any competing fluorescence process. In impulsive coherent Raman spectroscopy, the same spectral information is contained in the two interferograms produced by the intensity modulation of the filtered red- or blue-shifted radiation. As it is shown in Fig. 1, and already partly explained above, the blue- and red-shifts alternate, i.e. the Stokes and anti-Stokes modulation signals have the opposite phase. It is thus possible to simultaneously monitor at the output of the sample the intensity modulation experienced by the Stokes and anti-Stokes radiations on two independent detectors. The difference of the two detector signals enhances the interferometric signal intensity and cancels the common amplitude noise.

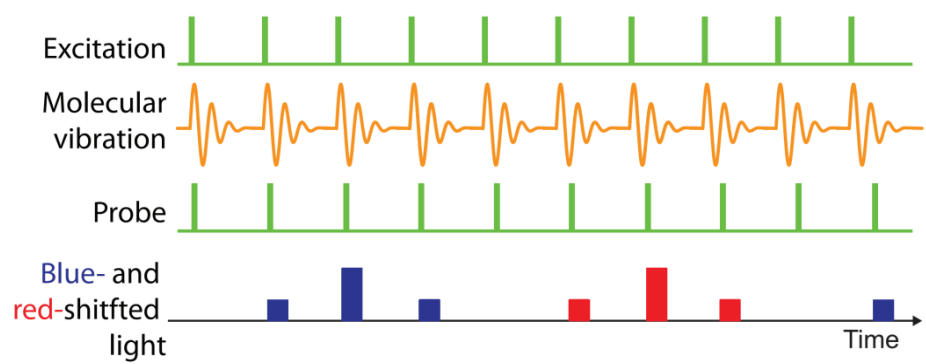

Figure1: Principle of impulsive coherent Raman spectroscopy. For simplicity, only one vibrational level is assumed to lie within the excitation bandwidth. The excitation pulse periodically drives by a two-photon process the vibrational transition. The second pulse probes, with a linearly increasing time-delay, the molecular vibration amplitude. This generates a modulation in the alternatively-generated blue- and red-shifted light. We monitor the amount of blue- and red-shifted light on two photodetectors.

Figure 2 sketches the experimental set-up that we use for a proof-of-principle demonstration. A Ti:Sa modelocked laser (Synergy-20UHP, Femtolasers) emits a train of 20-fs pulses with a repetition frequency of 100 MHz and an average power of 1.1 W. Its central wavelength is 795 nm (12580 cm$^{-1}$). A scanning Michelson interferometer generates, from the output of the Ti:Sa laser, pairs of pulses with a linearly increasing (or decreasing) time delay. The home-made Michelson interferometer consists of a pellicle beam splitter and two hollow cube corners, one being mounted on a motorized linear translation stage. The beam of a narrow-linewidth continuous wave ytterbium-doped fiber laser (Rock, NP Photonics) at 1040 nm (9615 cm$^{-1}$) is passed through the Michelson interferometer together with the Ti:Sa beam. The interferogram of such monochromatic source is sinusoidal and its zero-crossings serve to sample the Raman interferometric signal at evenly-spaced optical retardations. We set the speed of the linear stage to 5 cm/s and the clock frequency is thus around 96 kHz. At the output of the Michelson interferometer, on the beam path of the Ti:Sa laser, a pair of chirped mirrors (Layertec) and two fused silica wedges pre-compensate for the second order dispersion of the optics coarsely and finely, respectively. A combination of longpass and shortpass filters before the sample limits the spectral bandwidth of the pulses to 750-835 nm (11976-13333 cm$^{-1}$) in order to improve rejection of the Ti:Sa beam in the detection of the red-shifted and blue-shifted signals. The pulses are focused, with a lens of 8-mm focal length, in the liquid sample, neat hexafluorobenzene, in a 1-mm long cuvette. The total incident average power at the sample is 260 mW. The light transmitted by the sample in the forward direction contains the Ti:Sa excitation light as well as the blue- and red-shifted radiation of interest. To collect the blue-shifted anti-Stokes radiation, a shortpass filter (cut-off wavelength: 750 nm - 13333 cm$^{-1}$) is placed in front of one of the two Si photodiodes of a 125-kHz-bandwidth auto-balanced photodetector (Nirvana, Newport). The beam reflected by the shortpass filter is filtered by a longpass filter (cut-off wavelength: 835 nm -11976 cm$^{-1}$) and focused onto the second photodiode of the balanced photodetector. Tilting the filters makes it possible to tune their cut-off wavelengths and to optimize the balance detection. The signal of the differential photodetector is electronically lowpass filtered and sampled at a rate determined by the clock frequency (96 kHz) by a 16-bit data acquisition board. The interferograms are Fourier transformed to reveal the spectra.

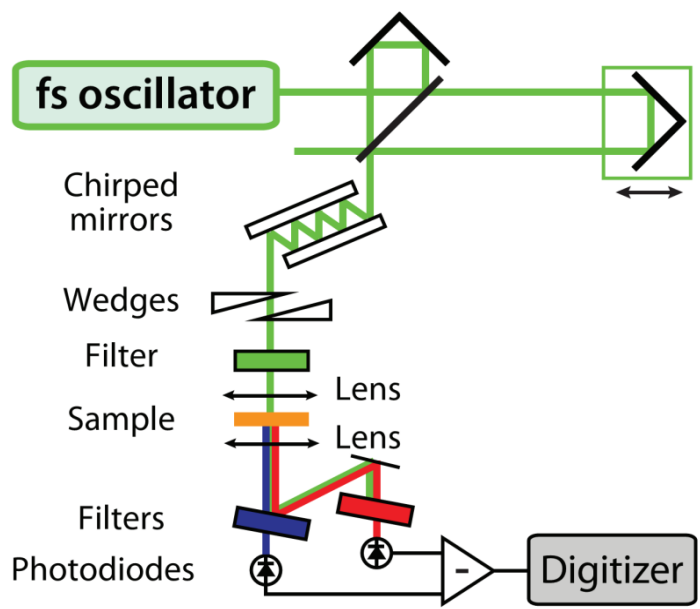

Figure 2: Experimental setup for differential coherent Raman Stokes and anti-Stokes spectroscopy. See text for details.

Proper dispersion compensation is a key to this experimental concept. Effective impulsive Raman excitation requires laser pulses that are shorter than a molecular vibration period. The two-photon Raman-like excitation is an instantaneous process. All pairs of frequency components of the excitation laser which have a difference frequency matching that of a vibrational level may contribute to the excitation. Therefore dispersion in the excitation pulse reduces the efficiency of excitation and the bandwidth of the Raman spectrum. For the detection of the molecular transitions, we measure an intensity modulation in the generated Stokes and anti-Stokes radiations after spectral filtering. A chirped probe pulse decreases the visibility of the modulation fringes. Furthermore, for differential detection, the generated coherent Raman radiation is isolated by means of spectral filtering. Essentially the red-end of the Stokes signal and the blue-end of the anti-Stokes signal are used. An unwanted phase shift between the two interferograms may thus occur. In practice, fine tuning of the phase-shift is achieved by simultaneously monitoring the Stokes and anti-Stokes interferograms and optimizing the amount of dispersion with the translation of wedges in the beam path before the sample.

The measured interferograms corresponding to the Stokes (Fig.3a), anti-Stokes (Fig.3b) and differential (Fig.3c) signals are displayed as a function of the optical delay between the excitation and probe pulses. The interferograms are measured within 4 s: they result from the averaging of 100 interferograms, each measured within 40 ms. The three interferograms are acquired sequentially, first by blocking one or the other photodiode and then by using the balanced detection. Each interferogram comprises 4096 samples. A maximum optical delay of 14 ps leads to an apodized spectral resolution of 4 cm$^{-1}$. As expected, similar interference patterns are observed in both Stokes (Fig.3a) and anti-Stokes (Fig. 3b) interferograms. When zooming into the traces, the opposite phase in the two interferograms appears conspicuously in the comparison of Figs. 3d and 3e. The differential interferogram is shown in Figs. 3c and 3f. The enhancement of signal to noise ratio is already obvious in the comparison of the interferograms: the differential interferogram exhibits nicer interferometric modulation for optical delays shorter than 8 ps and is exempt of the low frequency noise that dominates the Stokes and anti-Stokes interferograms for optical delays longer than 10ps.

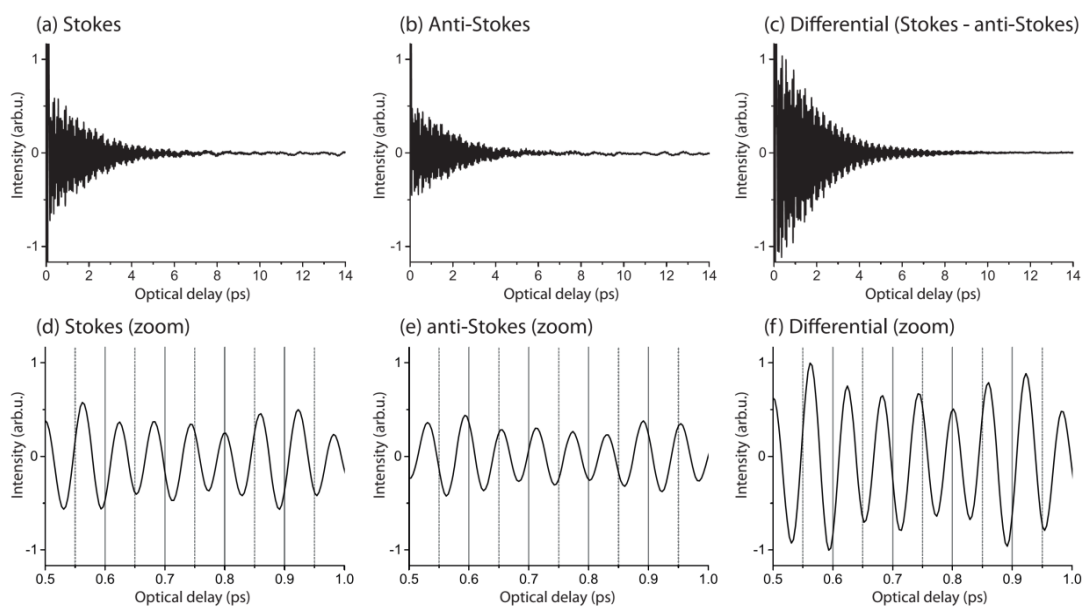

Figure 3: Interferograms of the (a) Stokes signal (CSRS), (b) anti-Stokes signal (CARS), (c) differential signal of Stokes and anti-Stokes signals. (d)-(f) zoom into a part of the interferograms (a)-(c).

Figure 4 shows the spectra after Fourier transformation of the interferograms of Fig. 3. The Stokes (Figs. 4a, 4d), anti-Stokes (Figs. 4b, 4e) and differential (Figs. 4c, 4f) Raman spectra are displayed both on linear and logarithmic scales. Around an optical delay of zero, the nonresonant four-wave mixing signal resulting from the interference between the overlapping excitation and probe pulses produces a strong signal. To suppress the interferometric non-resonant contribution, we Fourier-transform a time-windowed portion of the interferograms, which excludes the small optical delays. Our spectra are thus non-resonant background free, as already demonstrated in other impulsive coherent Raman spectroscopy implementations. The spectra span over about 1300 cm$^{-1}$ with a resolution of 4 cm$^{-1}$ using triangular apodization. The spectral span of Raman shifts is limited by the spectral bandwidth of the femtosecond laser and the resolution is set to be narrower than the intrinsic linewidth of the transitions of the hexafluorobenzene sample. In all spectra, the three Raman bands $\nu_{10}$, $\nu_8$, $\nu_2$ of hexafluorobenzene [13] are observed at 369, 442 and 559 cm$^{-1}$, respectively. However the Stokes and anti-Stokes spectra exhibit low-frequency narrow-linewidth noise that obscures the 369-cm$^{-1}$ hexafluorobenzene band. Such artefacts are efficiently suppressed in the differential spectrum. The signal-to-noise ratio of the strongest hexafluorobenzene transition at 559 cm$^{-1}$ is determined as being the intensity of its signature divided by the root-mean-square (rms) noise in the 800-1000 cm$^{-1}$ region. It is found to be respectively 770, 470 and 1520 in the Stokes, anti-Stokes and differential spectra, respectively. Therefore the enhancement of signal-to-noise ratio due to differential detection is higher than 2. Interestingly, the rms value of the baseline in the 200-350 cm$^{-1}$ region, for a same intensity of the 559 cm$^{-1}$ band, (we arbitrarily set this rms value to 1 in the Stokes spectrum) is 1, 2.1 and 0.28, respectively in the Stokes, anti-Stokes and differential spectra, respectively. This shows that our differential measurement technique is particularly efficient in suppressing low-frequency common noise, which is mostly due to the femtosecond laser. This capability of noise cancelation may prove particularly important if femtosecond fiber lasers were used, as these usually exhibit a noise spectrum which is broader (and extending towards higher frequencies) than that of solid-state mode-locked lasers.

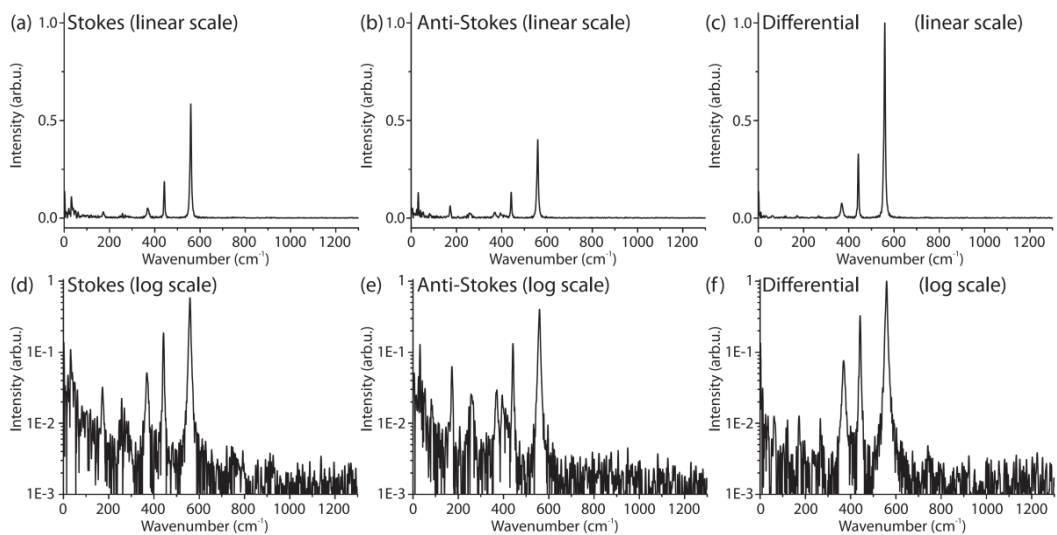

Figure 4: Raman spectra of the (a) Stokes signal (CSRS), (b) anti-Stokes signal (CARS), (c) differential spectrum between Stokes and anti-Stokes interferogram signals. (d)-(f) display (a)-(c) on a logarithmic intensity scale.

In conclusion, we have harnessed a yet-unexploited property of impulsive coherent Raman spectroscopy: the opposite phase of the generated Stokes and anti-Stokes interferograms renders differential detection possible. We have experimentally demonstrated an improvement of the signal-to-noise ratio of a high resolution broad spectrum measured by Michelson-based femtosecond coherent Raman spectroscopy. Raman hyperspectral imaging for microscopy applications currently attracts much interest and a variety of novel and powerful approaches are being explored [9,14–17]. In particular dual-comb coherent impulsive Raman spectroscopy holds much promise [9] for rapid and well-resolved hyperspectral imaging. Our technique may prove particularly useful for enhancing the sensitivity of such dual-comb experiments without sacrificing performance.

Support by the Max Planck Foundation, the European Laboratory for Frequency Comb Spectroscopy, the Munich Center for Advanced Photonics and the European Research Council (Advanced Investigator Grant 267854) are acknowledged.